\let\Xdocument\document
\documentclass{iau}
\let\document\Xdocument

\usepackage{amsmath}
\usepackage{graphicx}
\usepackage{multirow}
\usepackage{hyperref}

\begin{document}

\lefttitle{Shajib}
\righttitle{Strong lensing by galaxies}

\jnlPage{1}{7}
\jnlDoiYr{2023}
\volno{381}
\doival{10.1017/xxxxx}

\newcommand{\araa}{ARAA}
\newcommand{\apss}{APSS}
\newcommand{\nat}{Nature}
\newcommand{\ssr}{SSR}
\newcommand{\aapr}{AAPR}
\newcommand{\apj}{ApJ}
\newcommand{\apjs}{ApJS}
\newcommand{\apjl}{ApJL}
\newcommand{\mnras}{MNRAS}
\newcommand{\prl}{PRL}
\newcommand{\prd}{PRD}
\newcommand{\aap}{A\&A}
\newcommand{\jcap}{JCAP}
\newcommand{\pasj}{PASJ}
\newcommand{\aj}{AJ}

\aopheadtitle{Proceedings of IAU Symposium}
\editors{H.~Stacey, C.~Grillo, A.~Sonnenfeld, eds.}

\title{Strong lensing by galaxies: past highlights, current status, and future prospects}

\author{Anowar J. Shajib}
\affiliation{Department of Astronomy \& Astrophysics, The University of Chicago, Chicago, IL 60637, USA}
\affiliation{Kavli Institute for Cosmological Physics,
    University of Chicago, Chicago, IL 60637, USA}
\affiliation{NHFP Einstein Fellow, Email: \email{ajshajib@uchicago.edu}}

\begin{abstract}
Galaxy-scale strong lensing is a powerful tool in Astrophysics and Cosmology, enabling studies of massive galaxies' internal structure, their formation and evolution, stellar initial mass function, and cosmological parameters. In this conference proceeding, we highlight key findings from the past decade in astrophysical applications of strong lensing at the galaxy scale. We then briefly summarize the present status of discovery and analyses of new samples from recent or ongoing surveys. Finally, we offer insights into anticipated developments in the upcoming era of big data shaping the future of this field, thanks to the Rubin, \textit{Euclid}, and \textit{Roman} observatories.
\end{abstract}


\maketitle

\section{Introduction} \label{sec:intro}

Strong lensing by galaxies provides a valuable probe of the internal structure of the lensing galaxies \citep{Shajib22d}. Since elliptical galaxies are the most common type of lensing galaxies, studies of strong lensing galaxies have mainly focused on massive ellipticals. The internal structure of elliptical galaxies is shaped through cosmic time by the initial baryonic infall, star formation, subsequent outflow induced by baryonic feedback mechanisms, and hierarchical formation through mergers. All of these baryonic processes are also accompanied by adiabatic contraction and expansion of the dark matter. Thus, constraining the internal structure of galaxies, that is, the distribution of baryonic and dark matter, at different redshifts can shed light on their formation and evolutionary history. Stellar dynamics, especially the spatially resolved kind, of local elliptical galaxies have provided great insight into their formation and evolution \citep{Cappellari16}. However, strong lensing offers unique complementarities to stellar dynamics in providing more accessible and informative data for galaxies at redshift $z \gtrsim 0.5$ and by providing an independent tracer of the mass to break degeneracies inherent to both probes, when combined.

Since the first discovery of strong lensing systems at the galaxy scale in the eighties, the field has come a long way in efficiently harnessing the rich information in large samples of strong lenses. Thanks to these large samples, for example, the Sloan Lens ACS Survey \citep[SLACS;][]{Bolton06} and the Cosmic Lens All Sky Survey \citep[CLASS;][]{Myers95}, the internal structure of elliptical galaxies at $z\sim0.3$ were unveiled with significant precision in the 2000s. However, at the beginning of the past decade (2010s), \citet{Treu10b} posed the following two main outstanding questions to be solved by strong lensing studies of galaxies:

\begin{enumerate}
	\item How do luminous and dark matter density profiles evolve over cosmic time?
	\item Does the dark matter density profile universally follow the Navarro--Frenk--White \citep[NFW;][]{Navarro96, Navarro97} profile predicted by cosmological simulations?
\end{enumerate} 

In this proceeding, we briefly introduce the progress made on these two questions over the past decade (Section \ref{sec:past}). We then introduce the current status of the discovery of larger lens samples and their analyses (Section \ref{sec:present}), before concluding the proceeding with a discussion on the prospects for the forthcoming era of big data (Section \ref{sec:future}). The readers are invited to see \cite{Shajib22d} for a more comprehensive review of the topic.

\section{Highlights of past results} \label{sec:past}

In this section, we present some major results from the literature on the internal structure of elliptical galaxies (Section \ref{sec:structure}) and the stellar initial mass function (IMF; Section \ref{sec:imf}).

\subsection{Internal structure of elliptical galaxies} \label{sec:structure}

Numerous strong lensing studies found that the total density profile in elliptical galaxies is well approximated by a power-law profile, that is, $\rho(r) \propto r^{-\gamma}$. These studies also showed that the power-law profile is nearly isothermal, that is, $\gamma \sim 2$. This result holds for the sample-mean of the logarithmic slope $\gamma$ constrained by both joint lensing--dynamics analyses [e.g., $2.08 \pm 0.03$ from the SLACS \citep{Auger09}, $2.11 \pm 0.02$ from the BOSS Emission-Line Lens Survey  \citep[BELLS;][]{Bolton12}, $2.05 \pm 0.06$ from the Strong Lensing Legacy Survey \citep[SL2S;][]{Sonnenfeld13}] and strong-lensing-only analyses [e.g., $2.08 \pm 0.03$ from SLACS \citep{Shajib20}, and $2.09 \pm 0.03$ from SLACS and BELLS \citep{Etherington22}]. These analyses also found intrinsic scatters in $\gamma$ between $0.13-0.19$.

\begin{figure*}
	\includegraphics[width=\textwidth]{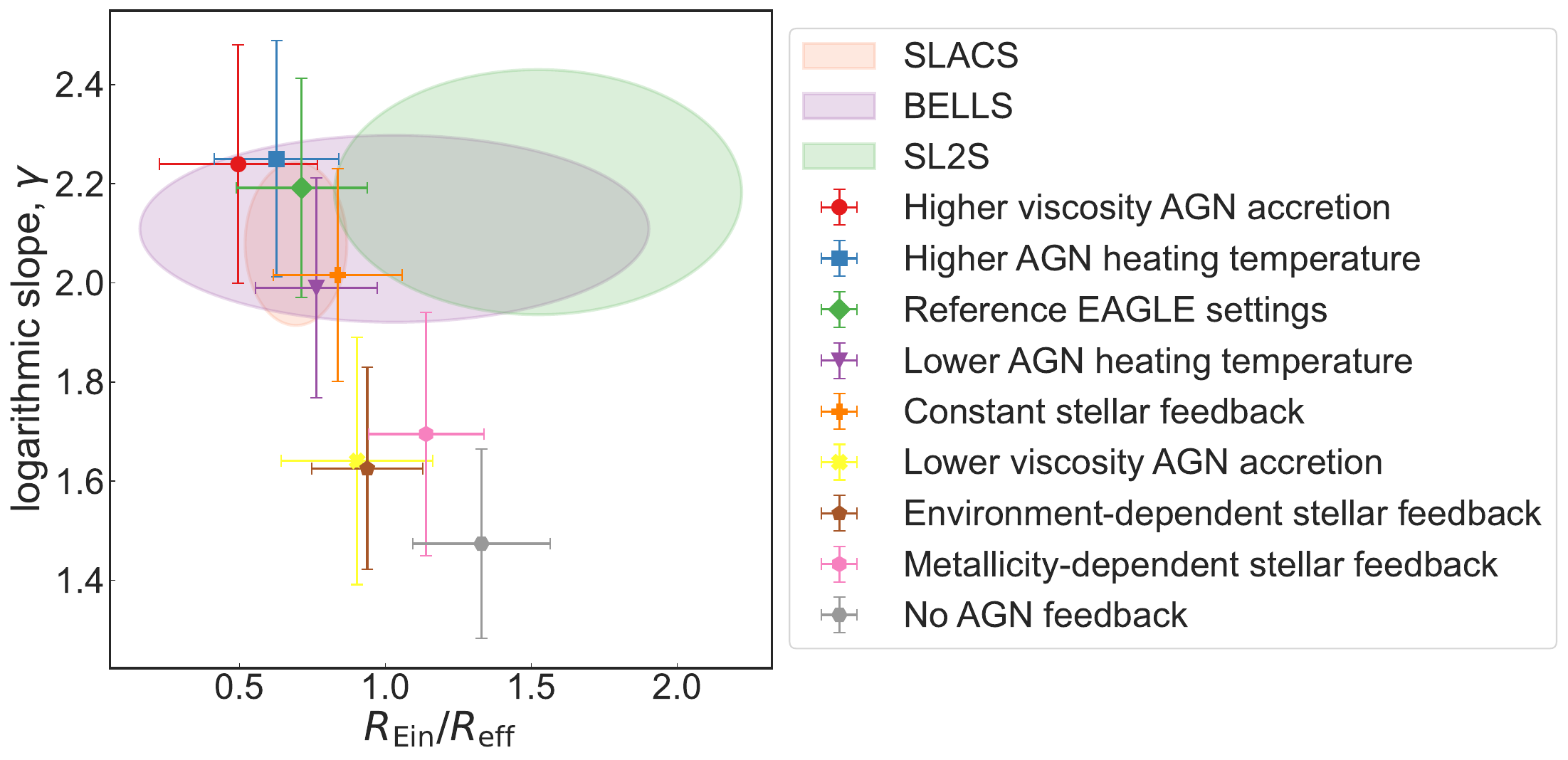}
	\caption{Comparison of the constrained logarithmic slope $\gamma$ and Einstein radius $R_{\rm Ein}$ between observations (shaded ellipses) and predictions from the EAGLE simulation with various baryonic physics prescriptions (points with errorbars). Some prescriptions, such as the no-AGN feedback model (grey point), can be ruled out. Figure re-illustrated from \citet{Mukherjee21}.}
	\label{fig:seagle}	
\end{figure*}

The logarithmic slope $\gamma$ constrained by the aforementioned studies can be compared to those from cosmological hydrodynamical simulations with various prescriptions of baryonic physics to learn about the particular baryonic physics that has been at play in shaping the elliptical galaxies. For example, \citet{Mukherjee21} rule out scenarios such as no AGN feedback and, lower-viscosity AGN accretion, or environment-dependent stellar feedback (Fig. \ref{fig:seagle}).

\begin{figure*}
	\centering
	\includegraphics[width=0.75\textwidth]{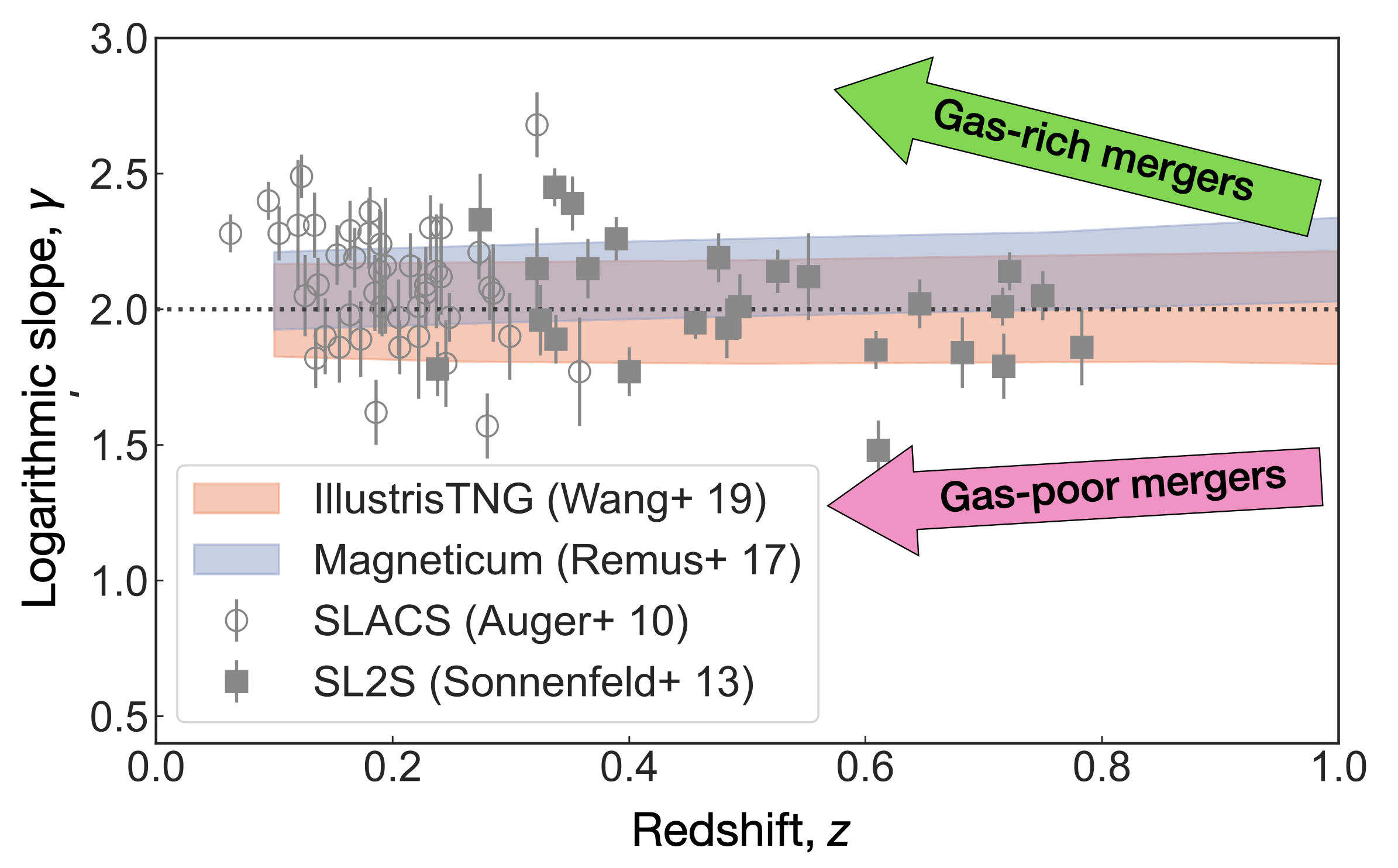}
	\caption{Evolution of the average logarithmic slope $\gamma$ with redshift $z$. The green and pink arrows indicate the direction of evolution in $\gamma$ for gas-rich and gas-poor mergers, respectively. Simulations (orange and blue shaded stripes) indicate gas-poor mergers make the slope slightly shallower with decreasing redshift. However, joint lensing--dynamics observations indicate the opposite trend favoring the gas-rich merger scenario.}
	\label{fig:gamma_vs_z}	
\end{figure*}

The evolution of the slope of the mass density profile over redshift can additionally provide further insights into the baryonic feedback processes and merger types that played crucial roles during the lifespan of elliptical galaxies at $z < 1$. Hydrodynamic simulations, such as IllustrisTNG and Magneticum, show that gas-poor mergers leave the logarithmic slope in elliptical galaxies unchanged, or make it slightly shallower with decreasing redshift \citep[Fig. \ref{fig:gamma_vs_z};][]{Remus17, Wang20}. Conversely, gas-rich mergers would have made the average logarithmic slope steeper. In apparent contrast with the simulations, joint lensing--dynamics studies found that the logarithmic slope $\gamma$ gets steeper with decreasing redshift, thus favoring the gas-rich merger scenario \citep{Ruff11, Bolton12}. However, this discrepancy between simulations and observations can also be explained by systematic effects, such as selection effect \citep{Sonnenfeld15}, projection effect \citep{Remus17}, and parametrization of the stellar anisotropy \citep{Xu17}.

Furthermore, \citet{Etherington22} found no correlation in the logarithmic slopes obtained from lensing-only analysis and joint lensing--dynamic analysis. These authors, therefore, argue that the true mass distribution in elliptical galaxies must deviate from a pure power-law form. To allow more radial flexibility than a power law, several studies modeled the mass distribution in elliptical galaxies using a two-component model with one mass profile describing the dark matter halo (typically with an NFW profile) and the other describing the stellar mass distribution. However, some disagreement exists in the literature about the departure from the ``vanilla'' (i.e., not contracted nor expanded) NFW profile and the presence of a mass-to-light ratio gradient in the stellar mass distribution. \citet{Dutton14} found the vanilla NFW profile fits the lensing and dynamics data, whereas scenarios with adiabatic contraction or expansion do not. However, \citet{Oldham18} found a bimodality in their sample with the inner slope shallower (expanded) than the vanilla NFW in one mode and steeper (contracted) in the other. In contrast, \citet{Sonnenfeld18} found that a mass-to-light ratio gradient in the stellar mass profile, not a departure from the vanilla NFW profile, fits better the combined dataset from strong lensing, dynamics, and weak lensing. Finally, \citet{Shajib21} ruled our contraction or expansion in the dark matter halos and found no strong evidence in favor of a mass-to-light ratio gradient in the stellar mass from the combination of strong lensing, dynamic, and weak lensing data. However, these discrepancies in the literature could potentially arise from different treatment stellar anisotropy parameterizations and light profile assumptions in the dynamical modeling.

\subsection{Stellar IMF} \label{sec:imf}

\begin{figure*}
	\centering
	\includegraphics[width=0.7\textwidth]{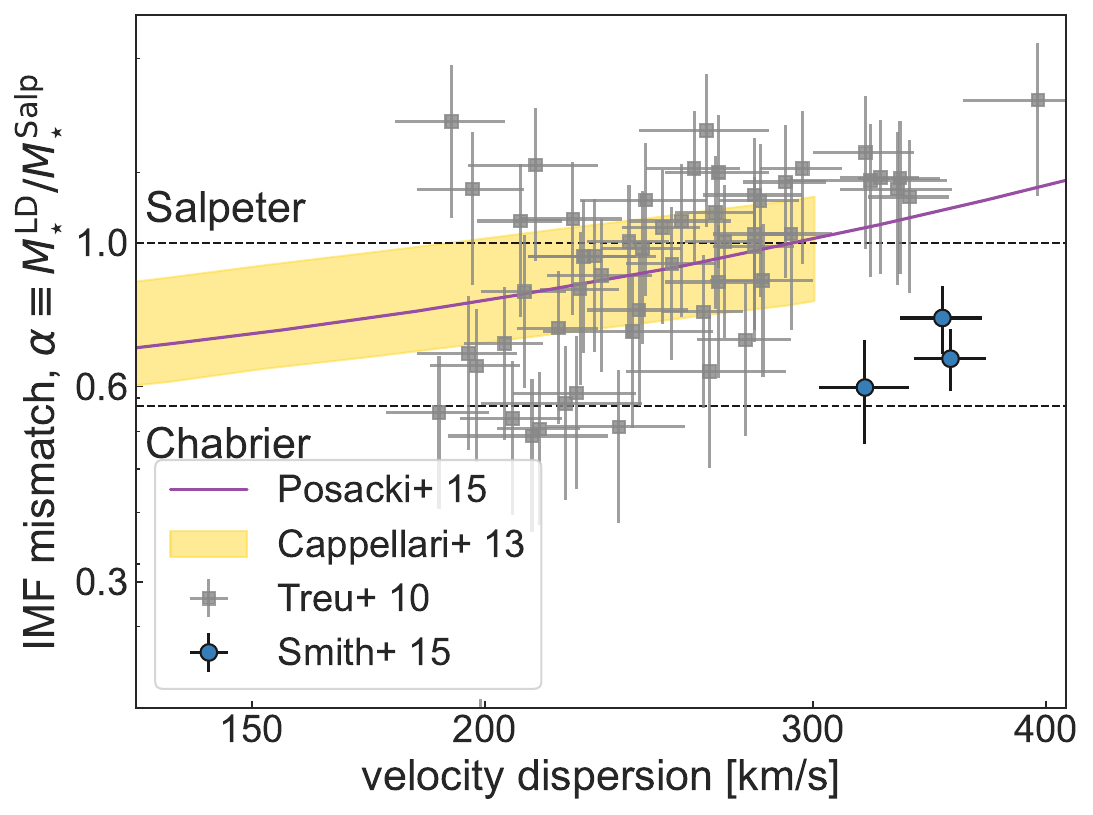}
	\caption{Compilation of results on the stellar IMF from lensing and dynamics. The IMF mismatch parameter $\alpha = 1$ represents the Salpeter IMF. The majority of constraints from lensing and dynamics indicate a trend in the IMF with the velocity dispersion \citep[purple line,][]{Posacki15}. However, some measurements, for example, the blue points from \citet{Smith15}, disagree with this trend. Figure re-illustrated from \citet{Smith15}.}
	\label{fig:imf}	
\end{figure*}

Most strong lensing studies found the stellar IMF to be heavier \citep[i.e., Salpeter IMF; e.g.,][]{Treu10, Spiniello11, Sonnenfeld12} than that found in the Milky Way (i.e., Chabrier IMF). These results also agree very well with several non-lensing constraints on the IMF, such as those based on stellar dynamics and the stellar population synthesis method \citep{Cappellari12, LaBarbera13, Spiniello14}. Fig. \ref{fig:imf} illustrates this agreement between lensing-based and dynamics-based studies and the apparent dependency of the IMF on the velocity dispersion \citep{Posacki15}. 

However, there are also reports of a lighter IMF, such as the Chabrier IMF, from lensing-based studies \citep[e.g.,][]{Ferreras10, Smith15, Sonnenfeld19}. However, systematics stemming from parameterizations of the dark matter profile, the mass-to-light-ratio gradient, or the stellar anisotropy cannot be ruled out as the source of this disagreement. 

Despite much progress with multiple large samples of lenses over the last decade, the two questions posed in Section \ref{sec:intro} are yet to be answered definitively. To achieve that goal in the future, larger statistical samples, combining strong lensing, weak lensing, and dynamics, and adopting models with more radial flexibility than the power law would be essential. In the next section, we describe the current efforts in these aspects.

\section{Presently emerging samples and analyses} \label{sec:present}

One way to achieve a larger statistical sample than the previously available ones is to create a ``super-sample'' assembled from archival samples. Following this way, Project Dinos (PI: Shajib) was created to study the evolution of elliptical galaxies' internal structure at $z < 1$ using as many suitable archival samples as possible. From this project, Tan et al. (in preparation) will present an analysis of the ``super-sample'' (SLACS$+$SL2S$+$BELLS) with available archival \textit{Hubble Space Telescope} (\textit{HST}) data and ground-based velocity dispersion measurements.

Recently, lots of new samples have also been discovered, primarily using machine-learning techniques applied on several ground-based surveys \citep[e.g.,][]{Agnello18, Delchambre19, Wong22, Lemon23}. An emerging sample with high-resolution \textit{HST} imaging and high-quality ground-based velocity dispersion measurements is the Astro3D Galaxy Evolution with Lenses (AGEL) sample \citep{Tran22}. This sample primarily consists of newly discovered strong lenses from the Dark Energy Survey data \citep{Jacobs19, Jacobs19b}, and includes mostly galaxy-scale lenses with a subset of group-scale lenses.

Recently, two \textit{HST} Schedule Gap Program has been approved to take imaging data of the newly discovered lens systems from different surveys: one for galaxy--galaxy lenses (HST-SNAP-17307, PIs: Tran \& Shajib), and the other for lensed quasar systems (HST-SNAP-17308, PI: Lemon). These two programs, combined, will obtain high-resolution \textit{HST} imaging of $\sim$450--600 galaxy-scale lenses over the next 3--5 years.

\section{Future prospects in the era of big data} \label{sec:future}

With the imminent launch of the Vera C. Rubin Observatory's Legacy Survey of Space and Time (LSST), and from current and future space-based surveys such as the \textit{Euclid} and the \textit{Roman Space Telescope}, even larger samples with size $\mathcal{O}(10^4)-\mathcal{O}(10^5)$ will be discovered \citep{Oguri10, Collett15}. Discovering these lenses would primarily utilize machine learning techniques with more novel algorithms currently being developed \citep[e.g.,][]{Akhazhanov22}. However, analyzing such unprecedentedly large samples will also pose a computational challenge. One way to tackle this challenge is to use machine learning for lensing parameter extraction from the data, which is currently an active area of research \citep[e.g.,][]{Hezaveh17, Morningstar19, Schuldt21, Adam22, Poh22, Sharma22, Biggio23}. However, depending on specific scientific requirements, the conventional forward modeling approach would still be favorable for a subset of all the new lenses. This subset of lenses will still be too large (e.g., $\mathcal{O}(10^3)$) for feasible analysis with the traditional method that requires a considerable amount of fine-tuning by a human modeler on a case-by-case basis. Several efforts are underway to develop automated lens modeling pipelines, e.g., \textsc{PyAutoLens} \citep{Nightingale21}, using \textsc{lenstronomy} \citep{Birrer18, Birrer21b, Shajib19, Shajib21, Schmidt22}, and using \textsc{Glee} \citep{Ertl23}.

\begin{figure}
    \centering
    \includegraphics[width=0.9\textwidth]{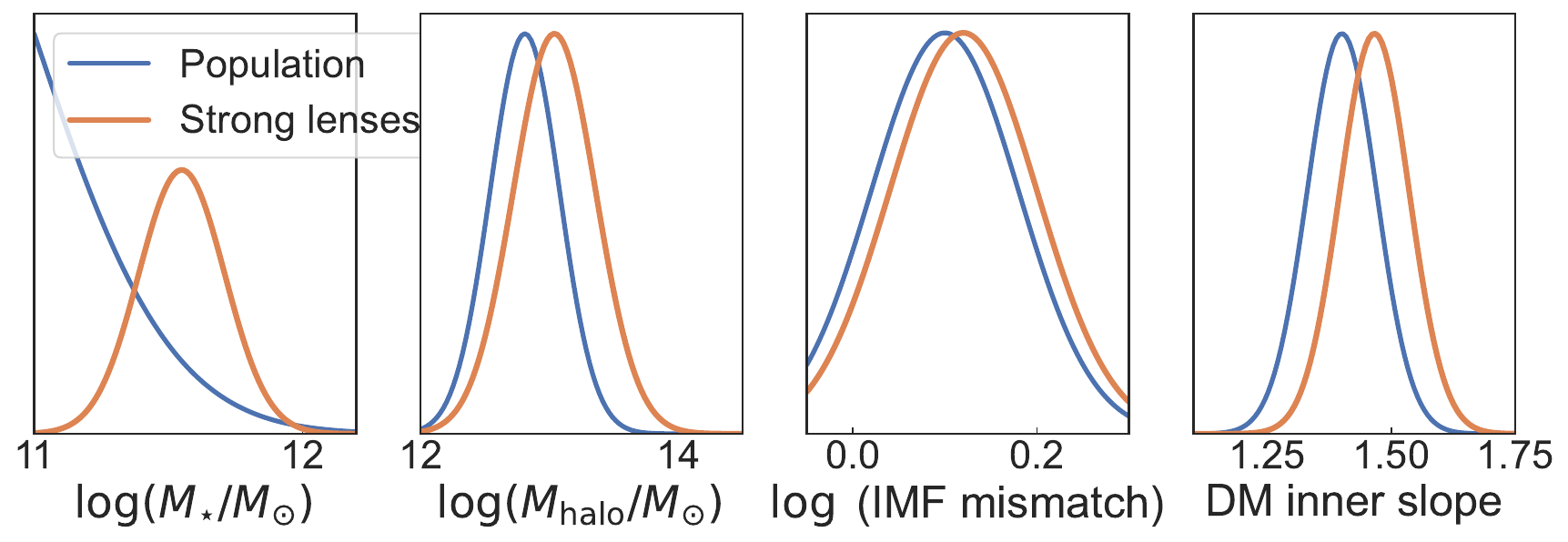}
    \caption{Impact of strong lensing selection on various properties of galaxies estimated by \citet{Sonnenfeld23}: from left to right, stellar mass $M_{\star}$, halo mass $M_{\rm halo}$, IMF mismatch $\alpha$, and dark matter (DM) inner slope. The IMF mismatch parameter in strong lensing galaxies is biased by 10\% from the parent population, and the dark matter inner slope is biased by 5\%.}
    \label{fig:selection}
\end{figure}

As the statistical precision grows tighter with these future large samples, systematic effects such as the selection function will become increasingly important, if not already. Using simulation, \citet{Sonnenfeld23} show that strong lensing samples can be biased in the IMF mismatch parameter by $\sim$10\% and in the dark matter inner slope by $\sim$5\% (Fig. \ref{fig:selection}). However, the selection effect from more factors, such as the halo concentration, environment, viewing angles, etc., could still be investigated by future studies. Accurately estimating the selection bias from these factors would allow future analyses of strong lens samples to correct the potential selection biases.

Although the last decade saw a large amount of progress in discovering new lens samples and in novel modeling and analysis techniques, several key questions about the internal structure and evolution of massive elliptical galaxies still lack definitive answers. This decade is expected to revolutionize all applications of strong lensing at the galaxy scale due to the two or more orders of magnitude increase in the lens sample size. Thus, Rubin, \textit{Euclid}, and \textit{Roman} observatories will play key roles in providing those answers in the forthcoming era of big data.

\begin{acknowledgements}
AJS was supported by NASA through the NASA Hubble Fellowship grant HST-HF2-51492 awarded by the Space Telescope Science Institute, which is operated by the Association of Universities for Research in Astronomy, Inc., for NASA, under contract NAS5-26555.
\end{acknowledgements}

\newcommand{\etal}{\textit{et. al}}

%
%


\end{document}